\documentclass[aps,prb,reprint,superscriptaddress,longbibliography]{revtex4-2}
\usepackage{amsmath,amssymb,bm,graphicx,booktabs}
\usepackage[colorlinks=true,citecolor=blue,linkcolor=blue,urlcolor=blue]{hyperref}
\newcommand{\I}{\openone}
\newcommand{\Li}{\mathcal{L}}
\newcommand{\Heff}{H_{\mathrm{eff}}}
\newcommand{\ket}[1]{\lvert#1\rangle}
\newcommand{\bra}[1]{\langle#1\rvert}

\newcommand{\rank}{\operatorname{rank}}
\newcommand{\Tr}{\operatorname{Tr}}

\newcommand{\chol}{\operatorname{chol}}
\newcommand{\OSR}{\operatorname{OSR}}

\begin{document}
\title{Dissipation-induced integrability in a spin-1 chain and a Bose--Hubbard mixture}
\author{Kaixiang Lu}\affiliation{School of Physics, Nankai University, Tianjin 300071, China}
\author{Mingdi Xu}\affiliation{School of Physics, Nankai University, Tianjin 300071, China}
\author{Xiang-Ping Jiang}\email{2015iopjxp@gmail.com}\affiliation{School of Physics, Hangzhou Normal University, Hangzhou, Zhejiang 311121, China}
\author{Lei Pan}\email{panlei@nankai.edu.cn}\affiliation{School of Physics, Nankai University, Tianjin 300071, China}
\date{November 9, 2026}
\begin{abstract}
We construct a nonintegrable spin-1 chain that acquires a Yang--Baxter
integrable no-jump Hamiltonian through a single dissipative channel per
bond. Energy is the only nontrivial finite-range charge of the isolated
periodic bulk. Interference
cancels the reverse spin conversions, leaving directed moves that obey
nil-Coxeter relations and generate commuting charges of increasing range
on the open chain. We obtain the full Liouvillian spectrum at any chain
length and identify an exponentially large set of blocked spin configurations
whose superpositions are all stationary. Resolving the bond outputs leaves
the no-jump Hamiltonian unchanged; each bond on which two configurations
differ then contributes equally to their dephasing. In a two-component
Bose--Hubbard chain, opposite loss phases produce counterpropagating
species whose motion and collisions
enter a common transfer matrix when interspecies attraction matches
intraspecies repulsion. At this coupling, redistributing a pair costs no
interaction energy in the atomic limit. Such bond interference offers a
route to integrability in other open quantum many-body systems.
\end{abstract}
\maketitle
\section{Introduction}\label{sec:introduction}
In an isolated integrable chain, conserved quantities leave a measurable memory of the initial state. After a quench in the XXZ chain, even quasi-local charges enter the generalized Gibbs ensemble that fixes the local observables~\cite{n_ilievski2015,n_ilievski2016}. Integrability is generally fragile against external perturbations or noise. Recently, however, it has been shown that coupling a chain to a reservoir can also generate a commuting family. A particularly interesting route starts with the one-dimensional Bose--Hubbard model. Zheng \textit{et al.} obtained its Bethe-ansatz solution and identified the Mott transition when bosons hop in only one direction~\cite{b5}. Ekman and Bergholtz showed how to reach this limit via bond loss~\cite{b6}: at a matched rate, loss through a common channel cancels the coherent hopping amplitude in one direction. Each jump also removes a particle, so the spectrum of the full generator follows from those of the conditional Hamiltonians at successive particle numbers.

This raises the question of how far this mechanism extends to a spin chain. A local spin conversion can select a direction on a bond while leaving exactly one spin on every site. The particle-number ordering that underlies boson loss is then unavailable. Even if the directed Hamiltonian has a Yang--Baxter family of commuting operators~\cite{b1,b2}, a jump can return population and coherence to states that the conditional evolution has already depleted~\cite{n_dalibard1992}. For some spin chains, the space of density operators splits into invariant sectors with integrable dynamics~\cite{n_essler2020}. The Bethe solution of one such XXZ sector shows how a boundary link controls the localization of decay modes at the chain ends~\cite{n_mao2024}. A different route imposes the Yang--Baxter equation directly on the Lindblad generator~\cite{b4}. Here we ask how a single bond reservoir affects a nonintegrable spin interaction and whether a coherent superposition survives the resulting jumps.

We consider a spin-1 parent Hamiltonian whose periodic bulk has no independent finite-range conserved charges, as follows from an exact local criterion~\cite{b13}. Each active bond couples four configurations to two oriented output states in a reciprocal way. One correlated jump cancels the reverse amplitudes. The surviving conversions obey nil-Coxeter relations and generate commuting operators up to the length of the open chain. Keeping the jumps yields a second result: the full relaxation spectrum can be read off from a triangular Liouvillian. The stationary configurations obey a simple local rule: every bond contains at least one spin in $\ket2$, i.e., outside the two input levels. The jump on a given bond then acts with the same amplitude on every allowed configuration. The reservoir preserves relative phases across this exponentially large set, as expected for a decoherence-free subspace~\cite{n_zanardi1997,n_lidar1998,n_albert2014}. Resolving the bond outputs causes the same phases to decay at a rate set by the number of distinguishable bonds. The two reservoirs thus have identical no-jump dynamics and different stationary states: a superposition of two blocked configurations can preserve its phase indefinitely in the first and relax to a mixture of the same configurations in the second.

For bosons, the two directions of motion can be assigned to different internal states. Sending each component to a separate reservoir lets one reverse the loss phase for one species while keeping their onsite collision energy fixed. There is a particular balance of repulsion within each species and attraction between species, $U_{aa}=U_{bb}=-U_{ab}$, at which the two counterpropagating components enter a common transfer matrix. At the same coupling, a pair containing one atom of each species can be redistributed among sites at zero energy cost in the atomic limit. Thus the interaction selected by integrability already has a direct signature in the onsite spectrum.

\begin{figure*}[t]
\includegraphics[width=\textwidth]{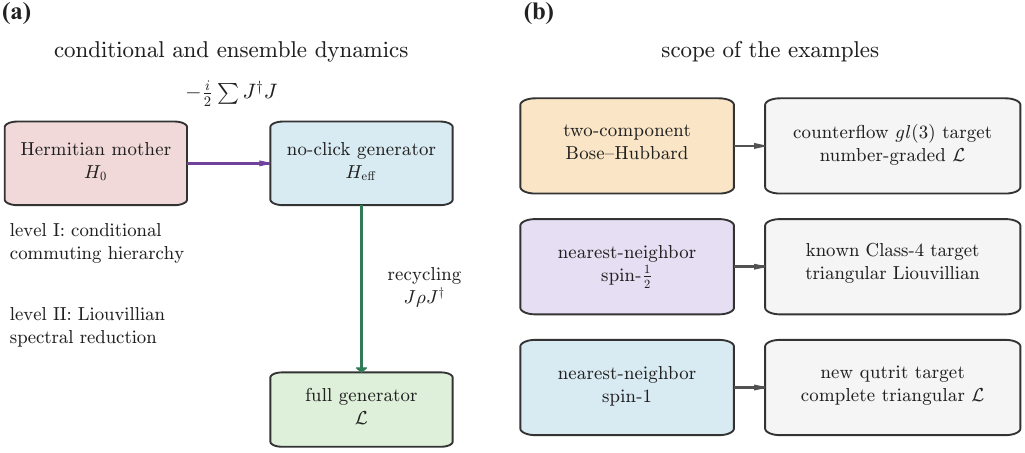}
\caption{Two roles of bond dissipation. (a) The loss kernel fixes the
no-click Hamiltonian for a specified monitoring scheme. Restoring the
recycling term gives the full density-matrix evolution.
(b) Conditional Yang--Baxter structures and the corresponding Liouvillian
spectral reductions for the bosonic mixture, the known spin-$1/2$
benchmark, and the spin-1 chain.}
\label{fig:1}
\end{figure*}

The rest of this paper is organized as follows. In Sec.~\ref{sec:framework}, we give the local dissipative construction (Fig.~\ref{fig:1}). Section~\ref{sec:spin1} develops the spin-1 chain and its relaxation spectrum. In Sec.~\ref{sec:bosons}, we derive the two-species model and describe a possible optical-lattice realization. Section~\ref{sec:conclusion} concludes.

\section{Dissipative construction}\label{sec:framework}
We consider a Hermitian Hamiltonian $H_0$ coupled to Markovian reservoirs by
jump operators $L_{j\mu}$ supported on at most two neighboring sites.
The density matrix obeys
\begin{equation}
\dot\rho=\Li\rho=-i[H_0,\rho]+\sum_{j,\mu}\mathcal D[L_{j\mu}]\rho,
\label{eq:1}
\end{equation}
where $\mathcal D[L]\rho=L\rho L^\dagger-\{L^\dagger L,\rho\}/2$.
For a trajectory conditioned on no detections in these specified output channels, propagation is generated by
\begin{equation}
\Heff=H_0-\frac i2\sum_{j,\mu}L_{j\mu}^\dagger L_{j\mu}.
\label{eq:4}
\end{equation}
The remaining recycling term restores trace-preserving evolution. With column
vectorization, the full generator is
\begin{equation}
\Li=-i(\I\otimes\Heff-\Heff^*\otimes\I)
+\sum_{j,\mu}L_{j\mu}^*\otimes L_{j\mu}.
\label{eq:5}
\end{equation}
It is this last term that distinguishes reservoirs with identical conditional
Hamiltonians. Its effect can already be seen in exact stationary states of
driven chains~\cite{n_prosen2011}. Resolving decay rates requires more:
quadratic generators admit a normal-mode decomposition~\cite{b3}, while
dephasing and two-body loss give interacting examples with Bethe-solvable
spectra~\cite{n_medvedyeva2016,n_nakagawa2021}. Collective spin jumps can
also yield an XXZ Richardson--Gaudin spectrum~\cite{n_claeys2022}.

For a finite-dimensional target density $h_\star$, its Hermitian part
$h_0=(h_\star+h_\star^\dagger)/2$ fixes the parent interaction.
Choose $\kappa$ so that the loss kernel $K=i(h_\star-h_\star^\dagger)+\kappa\I$
is positive semidefinite. A bond jump with $J^\dagger J=K$ then gives
$h_0-iJ^\dagger J/2=h_\star-i\kappa\I/2$.
The scalar shift sets the overall attenuation and leaves the conditional
commutators unchanged. Directionality comes from interference between the
coherent and dissipative matrix elements~\cite{n_metelmann2015,n_ashida2020}.
For the unbounded bosonic modes in Sec.~\ref{sec:bosons}, the corresponding
attenuation is proportional to particle number on the ring.

The loss kernel fixes the no-jump term for a specified monitoring scheme.
It admits several physically distinct recycling maps. For example, resolving
orthogonal final bond states preserves the sum of the kernels while changing
the coherences returned by a jump. This freedom is central to dissipative
state engineering~\cite{n_diehl2008,n_verstraete2009,n_kraus2008}.
Below, the conditional charges follow from bond algebra; the relaxation
spectrum follows by retaining recycling in Eq.~\eqref{eq:5}.

A simple benchmark is provided by the known spin-$1/2$ Class-4 density
$h_4$~\cite{b8,b9}. Its Hermitian part supplies reciprocal spin flips and
pair conversion, while a coherent bond jump cancels the reverse amplitudes.
An upper-triangular Cholesky factor of the positive loss kernel also makes
the full open-chain Liouvillian triangular. Appendix~\ref{app:qubit} gives
the explicit Hamiltonian, jump, bulk $R$ matrix, and relaxation spectrum.
\section{Spin-1 chain: integrability and stationary coherence}\label{sec:spin1}
\subsection{Reciprocal parent and correlated bond dissipation}
\begin{figure*}[t]
\includegraphics[width=\textwidth]{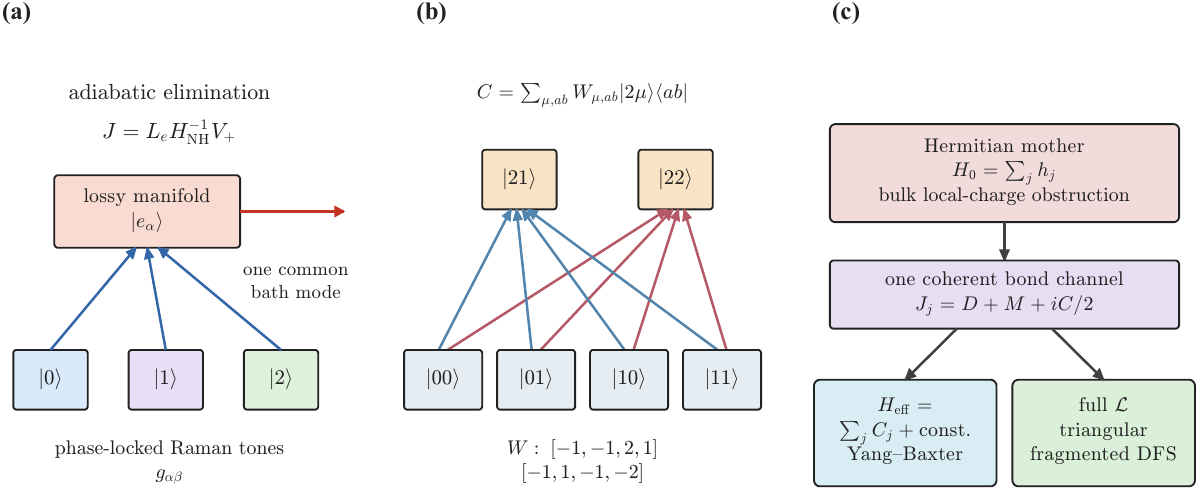}
\caption{Spin-1 conversion by coherent bond dissipation.
(a) Schematic auxiliary-manifold realization of one two-qutrit jump. The decay
paths share an indistinguishable output mode; phase-locked Raman amplitudes
set its matrix elements.
(b) The reciprocal parent coupling becomes the directed map $C$, with
nonzero block $W$.
(c) In the same local basis, the conditional Hamiltonian and the jump are
lower triangular, making the complete Liouvillian spectrum accessible.
The local-charge obstruction refers to the periodic bulk completion
(Appendix~\ref{app:D}).}
\label{fig:4}
\end{figure*}

Consider a chain of $L$ spins with open ends. We label the three states
$\ket0=\ket{m=-1}$, $\ket1=\ket{m=0}$, and $\ket2=\ket{m=+1}$, and order
the bond configurations lexicographically:
$\{\ket{00},\ket{01},\ket{02},\ket{10},\ket{11},\ket{12},
\ket{20},\ket{21},\ket{22}\}$. Before coupling to a reservoir the Hamiltonian is
\begin{align}
H_{1,0}&=\sum_{j=1}^{L-1}h_{j,j+1},\notag\\
h&=\frac12\ket{21}(-\bra{00}-\bra{01}+2\bra{10}+\bra{11})\notag\\
&\quad+\frac12\ket{22}(-\bra{00}+\bra{01}-\bra{10}-2\bra{11})
+\mathrm{H.c.}
\label{eq:36}
\end{align}
Each bond converts a pair in the $\{\ket0,\ket1\}$ manifold into
$\ket{21}$ or $\ket{22}$, and the Hermitian conjugate converts it back
[Fig.~\ref{fig:4}(b)]. The closed bulk is nonintegrable in the local-charge
sense. For the periodic completion $H_{1,0}^{\rm PBC}=\sum_{j=1}^{L}h_{j,j+1}$,
Hokkyo's criterion~\cite{b13} excludes strictly $k$-local conserved
quantities for $3\leq k\leq L/2$. Up to range two the conserved
space is spanned by the identity and the Hamiltonian. This covers any fixed
finite range once $L\geq 2k$ (Appendix~\ref{app:D}).

\begin{figure}[tb]
\centering
\includegraphics[width=\columnwidth]{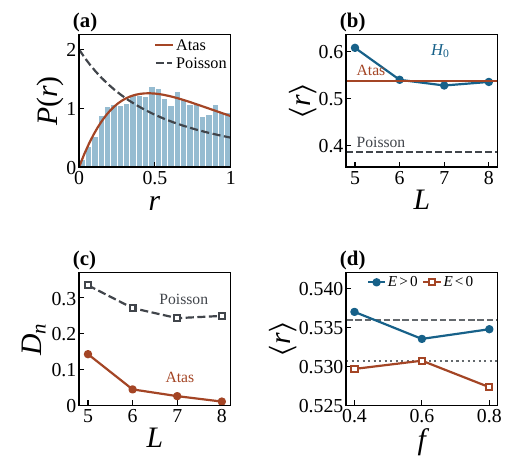}
\caption{Level statistics of the open spin-1 parent.
(a) Adjacent-gap ratios for the central 80\% of positive levels at $L=8$.
(b) Mean ratio and (c) CDF distances $D_n$ for $L=5$--8, with Atas and
Poisson references. (d) Positive- and negative-energy means at $L=8$
for three central window fractions $f$. Gray dashed and dotted lines mark
the Atas and large-GOE means, respectively. Numerical windows and
thresholds are given in Appendix~\ref{app:F}.}
\label{fig:5}
\end{figure}

The finite open chains used below show a corresponding spectral signature.
After removing exact zero modes, the adjacent levels in the connected
block repel. Their gap-ratio distribution approaches the Gaussian orthogonal
ensemble (GOE) prediction as $L$ increases [Fig.~\ref{fig:5}]. Windows on
both sides of zero give consistent results~\cite{b16,b17}. The symmetry
analysis and window dependence are given in Appendices~\ref{app:symmetry}
and~\ref{app:F}.

To obtain directional conversion, write $h=(C+C^\dagger)/2$, where
\begin{align}
C={}&\ket{21}(-\bra{00}-\bra{01}+2\bra{10}+\bra{11})\notag\\
&+\ket{22}(-\bra{00}+\bra{01}-\bra{10}-2\bra{11}).
\label{eq:42}
\end{align}
Its nonzero block is
$W=\bigl(\begin{smallmatrix}-1&-1&2&1\\-1&1&-1&-2\end{smallmatrix}\bigr)$.
The nonzero singular values of $W$ are $\sqrt{11}$ and $\sqrt3$. The
loss kernel $K=4\I+i(C-C^\dagger)$ is thus positive, with eigenvalues
$4\pm\sqrt{11}$, $4\pm\sqrt3$, and five copies of $4$.

We choose the lower-triangular factor $J$ satisfying $J^\dagger J=K$,
with positive diagonal. The jump on each bond is explicitly
\begin{equation}
J=D+M+\frac i2C,
\label{eq:47}
\end{equation}
where $D$ is diagonal in the bond basis. On a bond containing $\ket2$ its
diagonal entry is $d_{ab}=2$. The remaining entries are
$d_{00}=2\sqrt{13/15}$, $d_{01}=2\sqrt{5/7}$,
$d_{10}=\sqrt{105/11}/2$, and $d_{11}=\sqrt{11}/2$.
The mixing term $M$ has two nonzero rows. In row $10$, columns $00,01$
have amplitudes $(7,45)/(2\sqrt{1155})$; in row $11$, columns $00,01,10$
have amplitudes $(-1,3,-4)/(2\sqrt{11})$. All its other entries vanish.
The term $iC/2$ converts low-manifold pairs into $\ket{21}$ and
$\ket{22}$. All of these paths feed one output on each bond. Their amplitudes thus add
coherently in the recycling term $J_j\rho J_j^\dagger$.

One way to obtain a common output is to drive phase-locked Raman transitions
through an auxiliary manifold coupled to a single lossy mode
[Fig.~\ref{fig:4}(a)]. Adiabatic elimination then gives a jump proportional
to $J$~\cite{b18}; the required drives and separation of time scales are
specified in Appendix~\ref{app:implementation}. Reservoir engineering in ion
traps~\cite{b19} and proposals for cold atoms~\cite{b20} provide precedents
for the ingredients of this construction.

\subsection{Nil-Coxeter conditional dynamics}
Between counts of the bond output, the loss term contributes
$(C-C^\dagger)/2-2i\I$ to the reciprocal density $(C+C^\dagger)/2$.
The coefficient of $C^\dagger$ vanishes, leaving the open-chain conditional
Hamiltonian
\begin{equation}
\Heff=\sum_{j=1}^{L-1}C_{j,j+1}-2i(L-1)\I.
\label{eq:51}
\end{equation}
The energy unit is fixed by Eq.~\eqref{eq:36}; a scale $g>0$ can be restored
by $h\to gh$ and $J\to\sqrt gJ$.

Every action of $C$ puts the left spin of its bond in $\ket2$. This blocks
a second action on the same bond and one on its left neighbor. It follows
that
\begin{equation}
C^2=0,\qquad C_{12}C_{23}C_{12}=C_{23}C_{12}C_{23}=0.
\label{eq:52}
\end{equation}
These nil-Coxeter relations~\cite{b21,b27} admit the regular braided matrix
\begin{equation}
\check R(u)=\I+uC,\qquad \check R^{-1}(u)=\I-uC.
\label{eq:53}
\end{equation}
Expansion of the braided Yang--Baxter equation gives equal quadratic terms
on both sides; both cubic terms vanish by Eq.~\eqref{eq:52}~\cite{b22}.

The orientation of the bond moves also gives commuting charges directly
for the chain with open ends. Writing $C_j=C_{j,j+1}$, define
\begin{equation}
S_k=\sum_{a=1}^{L-k}C_{a+k-1}\cdots C_a,\qquad
[S_k,S_\ell]=0,\quad 1\leq k,\ell<L.
\label{eq:charges-main}
\end{equation}
Here $S_1$ is the nonscalar part of $\Heff$. The stronger relation
$C_jC_{j+1}=0$, together with commutativity of separated bonds, proves the
identity for every length. The $L-1$ operators are linearly independent;
Appendix~\ref{app:C} gives the cancellation argument and an explicit witness
for each range.

For the open chain, $\sum_j C_j$ is nonzero and strictly lower triangular.
The conditional spectrum therefore collapses to $-2i(L-1)$, while the
nonzero matrix elements of $S_k$ describe the conversion of $k$ consecutive
bonds in spatial order. Nilpotency excludes a Hermitian image of the
nonzero directed density under similarity.

The representation can also be distinguished from other directed qutrit
chains. Every nonscalar additive one-site symmetry of $C$ is
nondiagonalizable. The number-conserving eclectic chains~\cite{b23},
the displayed rank-one supersymmetric lifts~\cite{b24,b25}, and the
charge-conserving qutrit solutions compared in Appendix~\ref{app:comparison}
admit a nonscalar diagonalizable charge~\cite{b26}. This distinction is
invariant under a uniform invertible change of onsite basis.

\begin{figure*}[t]
\includegraphics[width=\textwidth]{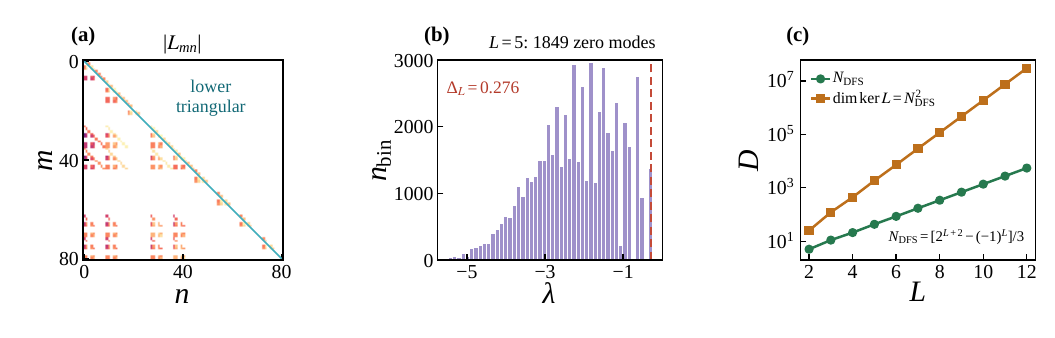}
\caption{Liouvillian spectrum and stationary manifold.
(a) Magnitude of the $81\times81$ single-bond superoperator; $m$ and $n$
index rows and columns in the product-operator basis.
(b) Eigenvalue counts $n_{\rm bin}$ from Eq.~\eqref{eq:60} at $L=5$.
The $1849=43^2$ zero modes are omitted; the first nonzero eigenvalue is
$-\Delta_{\Li}$.
(c) Dimensions $D$ of the DFS Hilbert space and its full stationary operator algebra.
A direct $L=3$ calculation gives $11^2$ zero modes, as predicted by
Eq.~\eqref{eq:66}.}
\label{fig:6}
\end{figure*}

\begin{figure}[tb]
\centering
\includegraphics[width=\columnwidth]{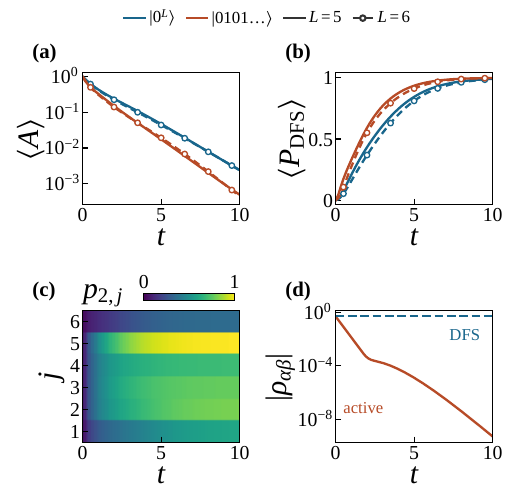}
\caption{Relaxation and coherence under the full Lindblad evolution.
(a) Active-bond density and (b) DFS weight for the all-zero and alternating
product states at $L=5,6$. (c) Site-resolved $\ket2$ population following a
quench from $\ket{0^6}$ at $L=6$. (d) Coherence of active and blocked cat
states at $L=5$. Only the blocked pair retains its initial magnitude $1/2$.}
\label{fig:7}
\end{figure}

\subsection{Full spectrum and coherent stationary manifold}
Strong local dissipation can separate decay modes into spectral bands in
the quantum Zeno limit~\cite{n_popkov2021}. In the spin-1 chain the full
spectrum is accessible at the finite rate where the reverse conversion
vanishes. The scalar shift in $\Heff$ gives the same
contribution $-4(L-1)$ on every diagonal entry. The recycling term
distinguishes different pairs of configurations. In the
lexicographic basis both $C$ and $J$ are lower triangular. The open-chain
embeddings preserve this order, so the no-jump term and the recycling term
in Eq.~\eqref{eq:5} are triangular in the same product-operator basis.
Reading off the diagonal gives every eigenvalue with its algebraic
multiplicity.
For strings $\bm a=(a_1,\ldots,a_L)$ and
$\bm b=(b_1,\ldots,b_L)$, the result is
\begin{equation}
\lambda_{\bm a,\bm b}=\sum_{j=1}^{L-1}
\left[d_{a_ja_{j+1}}d_{b_jb_{j+1}}-4\right].
\label{eq:60}
\end{equation}
The diagonal amplitudes $d_{ab}$ in Eq.~\eqref{eq:47} are real, positive,
and at most two, so all
eigenvalues are real and nonpositive. The dissipative gap for $L\geq2$ is
\begin{equation}
\Delta_{\Li}=4-2d_{00}.
\label{eq:63}
\end{equation}
It is attained by a string with a single active $00$ bond paired with an
inactive string. This gap fixes the slowest nonzero asymptotic exponential;
possible Jordan factors and spectral projectors still matter for a
mixing-time bound.

The stationary states have a direct interpretation. Let
$P_< =\ket0\bra0+\ket1\bra1$. A configuration $\ket{\bm a}$ is blocked
when every bond contains at least one $\ket2$, so that
$P_{<,j}P_{<,j+1}\ket{\bm a}=0$ for each $j$.
For every such string, $C_j\ket{\bm a}=0$ and
$J_j\ket{\bm a}=2\ket{\bm a}$. A count on any bond returns the same amplitude for
every allowed string. For $\ket{\bm a}\bra{\bm b}$ between two such
strings, recycling contributes $4\ket{\bm a}\bra{\bm b}$ per bond,
exactly canceling its conditional decay. Every superposition of allowed
strings thus belongs to a decoherence-free subspace
(DFS)~\cite{n_zanardi1997,n_lidar1998}.

Counting the allowed strings gives
\begin{equation}
N_{\rm DFS}(L)=\frac{2^{L+2}-(-1)^L}{3},\qquad
\dim\ker\Li=N_{\rm DFS}(L)^2.
\label{eq:66}
\end{equation}
Every summand in Eq.~\eqref{eq:60} vanishes only when both strings are locally
inactive. The displayed coherences therefore exhaust the algebraic zero
multiplicity; the zero eigenvalue is semisimple and no additional stationary
operators occur. Figure~\ref{fig:6} summarizes the decay spectrum and the
exponential growth of the coherent manifold.

\begin{figure*}[t]
\includegraphics[width=\textwidth]{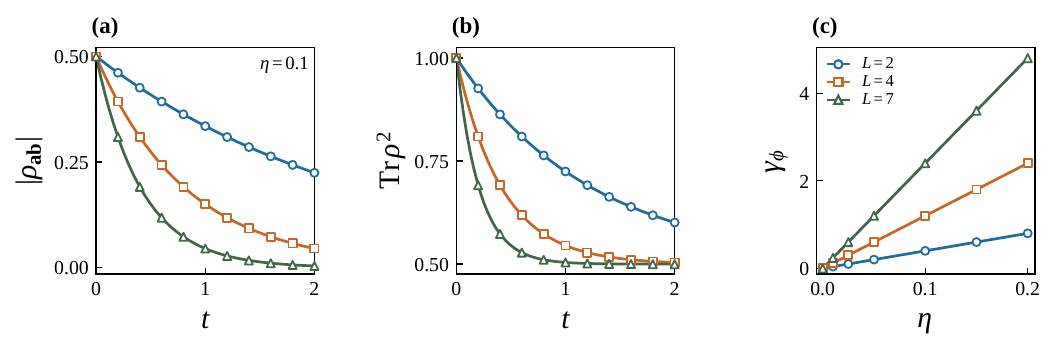}
\caption{Output distinguishability controls stationary coherence.
(a) Coherence and (b) purity at fixed output resolution $\eta=0.1$ for
$L=2,4,7$. The initial state is
$(\ket{\bm a}+i\ket{\bm b})/\sqrt2$, with $\bm a=2020\ldots$ and
$\bm b=2121\ldots$ truncated to $L$ sites. Every bond distinguishes the
two strings, giving $m=L-1$. Lines give Eq.~\eqref{eq:output-coherence};
symbols show propagation with the full generator.
(c) Decay rates fitted to the numerical coherences and the prediction
$\gamma_\phi=4\eta(L-1)$.}
\label{fig:8}
\end{figure*}

\subsection{Relaxation into blocked configurations}
We propagate the full Lindblad equation for open chains with $L=5,6$.
We measure the fraction of active bonds with the chain average
$A=(L-1)^{-1}\sum_{j=1}^{L-1}P_{<,j}P_{<,j+1}$ and the DFS weight with the projector
$P_{\rm DFS}=\sum_{\bm a\in{\rm DFS}}\ket{\bm a}\bra{\bm a}$.
The all-zero and alternating product states both begin with every bond
active. The dynamics creates blocking $\ket2$ occupations, reducing
$\langle A\rangle$ while transferring weight into the DFS
[Fig.~\ref{fig:7}(a),(b)]. The $L=6$ site-resolved population in
Fig.~\ref{fig:7}(c) develops a nonuniform pattern even though the jump is
repeated uniformly in the bulk. The spatial asymmetry reflects the orientation of the bond conversion.

The same relaxation distinguishes active and protected superpositions.
For the initial state $(\ket{00000}+i\ket{11111})/\sqrt2$, the branch
coherence decays. In contrast,
$(\ket{20202}+i\ket{21212})/\sqrt2$ lies entirely in the DFS and remains
stationary [Fig.~\ref{fig:7}(d)]. The coherence between these two blocked strings remains at its initial
magnitude $1/2$.

The direct $L=5,6$ comparison in Fig.~\ref{fig:7}(a),(b) shows the same blocking
mechanism at both sizes. Convergence and density-matrix checks are reported
in Appendix~\ref{app:F}.

\subsection{Environmental distinguishability destroys stationary coherence}
The coherent manifold relies on the common output of each bond jump.
To isolate this dependence, replace $J_j$ by
\begin{align}
J_{j,\mathrm c}&=\sqrt{1-\eta}\,J_j,\notag\\
J_{j,\alpha}&=\sqrt\eta\,P_{\alpha,j}J_j,\qquad
\sum_{\alpha=1}^{9}P_{\alpha,j}=\I,
\label{eq:output-jumps}
\end{align}
where $P_{\alpha,j}=\ket\alpha_j\bra\alpha$ resolves the local two-site output
and $0\leq\eta\leq1$. The sum of jump kernels remains
$\sum_\mu J_\mu^\dagger J_\mu=\sum_jJ_j^\dagger J_j$.
Both the Hermitian Hamiltonian and $\Heff$ are therefore independent of
$\eta$. Only the recycling map changes.

For allowed strings $\ket{\bm a},\ket{\bm b}$, a resolved bond contributes
to off-diagonal recycling only when their two local outputs coincide.
Writing $m$ for the number of bonds on which they differ, one obtains
\begin{equation}
\Li_\eta(\ket{\bm a}\bra{\bm b})
=-4\eta m\ket{\bm a}\bra{\bm b}.
\label{eq:output-rate}
\end{equation}
Thus the reservoir dephases the superposition at a rate set by its local
distinguishability. For two distinct allowed strings in the state
$(\ket{\bm a}+i\ket{\bm b})/\sqrt2$,
\begin{equation}
|\rho_{\bm a\bm b}(t)|=\frac12e^{-4\eta mt},\qquad
\Tr\rho(t)^2=\frac12+\frac12e^{-8\eta mt}.
\label{eq:output-coherence}
\end{equation}
Figure~\ref{fig:8} compares open chains of $L=2,4,7$ at the same
resolution. Extending the two alternating strings by one site adds one
distinguishable bond and increases the coherence decay rate by $4\eta$.

This change extends to the complete stationary operator space.
Each resolved jump remains triangular; Appendix~\ref{app:resolved} gives
the full spectrum and gap for arbitrary $\eta$. With rank-one resolution
on every bond of a connected chain, $L\geq2$,
\begin{align}
\ker\Li_0&=\mathcal B(\mathcal H_{\rm DFS}),\notag\\
\ker\Li_\eta&=\operatorname{span}\{\ket{\bm a}\bra{\bm a}:
\bm a\in{\rm DFS}\},\quad \eta>0.
\label{eq:resolved-kernel}
\end{align}
The stationary matrix algebra of dimension $N_{\rm DFS}^2$ becomes a
commutative population algebra of dimension $N_{\rm DFS}$. For any distinct
pair the coherence lifetime is $1/(4\eta m)$; it diverges continuously
as the output resolution is turned off.

Physically, the resolved reservoir carries a record of the local output
state. Equation~\eqref{eq:output-jumps} describes a family of different
unconditional generators with the same no-count evolution. Resolution on a
subset of bonds, or into coarser classes of outputs, can leave some pairs of
strings indistinguishable. Resolving each configuration on every bond
distinguishes all pairs and gives the diagonal stationary algebra in
Eq.~\eqref{eq:resolved-kernel} for each finite $L\geq2$.
\section{Counterpropagating bosons with onsite attraction}\label{sec:bosons}
Can two species move in opposite directions, collide, and still admit a
common transfer matrix? For a single species, coherent bond loss selects
the direction of hopping~\cite{b5,b6}. With two species the relative phases
of the loss amplitudes select their directions independently. The strength
of the interspecies interaction then enters the integrability condition.

\subsection{Reciprocal tunneling and coherent bond loss}
Let $a_j$ and $b_j$ annihilate bosons of the two species at site $j$ on a
ring, with $n_j^a=a_j^\dagger a_j$, $n_j^b=b_j^\dagger b_j$, and
$N_\sigma=\sum_j n_j^\sigma$. The Hermitian parent is
\begin{align}
H_{B,0}=-\frac t2\sum_j\big(&a_j^\dagger a_{j+1}+a_{j+1}^\dagger a_j\notag\\
&+b_j^\dagger b_{j+1}+b_{j+1}^\dagger b_j\big)+H_{\rm int},
\label{eq:10}
\end{align}
where $t,g>0$ and
$H_{\rm int}=(g/2)\sum_j[(n_j^a-n_j^b)^2-(n_j^a+n_j^b)]$.
The two intraspecies repulsions have strength $U_{aa}=U_{bb}=g$;
an $a$ boson and a $b$ boson on the same site gain the binding energy
$g$, i.e., $U_{ab}=-g$. In either fully polarized sector the parent
reduces to the conventional Bose--Hubbard chain, whose interacting
regime exhibits level repulsion~\cite{b11}.

Take one loss channel for each species on every bond:
\begin{equation}
J_{j,+}=\sqrt t(a_j-ia_{j+1}),\qquad
J_{j,-}=\sqrt t(b_j+ib_{j+1}).
\label{eq:11}
\end{equation}
The phase difference between the two ends of a bond is $-\pi/2$ for
$a$ and $+\pi/2$ for $b$. In the no-jump Hamiltonian, each loss kernel
cancels one of the two bare hopping amplitudes $-t/2$ and doubles the other.
On the ring, $\Heff=H_{\leftrightarrow}-it(N_a+N_b)$ with
\begin{align}
H_{\leftrightarrow}={}&-t\bigg[\sum_{j=1}^{L-1}
(a_j^\dagger a_{j+1}+b_{j+1}^\dagger b_j)\notag\\
&\qquad+a_L^\dagger a_1+b_1^\dagger b_L\bigg]+H_{\rm int}.
\label{eq:8}
\end{align}
An $a$ boson hops from $j+1$ to $j$ and a $b$ boson from $j$ to $j+1$,
both with amplitude $-t$. The onsite attraction $-g n_j^a n_j^b$ couples
these counterpropagating particles whenever they meet. The closing bond
continues the same motion through $1\to L$ for $a$ and $L\to1$ for $b$.
This is the periodic choice $\epsilon=1$ of Ref.~\cite{b5}, with spatial
reflection for the second species.

Each site belongs to two loss bonds on the ring. Cutting the closing bond
leaves the end sites with only one loss bond apiece. The shift added to
$H_{\leftrightarrow}^{\rm OBC}$ in the open conditional Hamiltonian is
then $-it(N_a+N_b)+it(n_1+n_L)/2$, where $n_j=n_j^a+n_j^b$.
It depends on the endpoint occupations even at fixed total particle number.
We use periodic boundaries for the transfer matrix and the finite-hopping
spectra. For comparison, in a continuum Bose gas with an imaginary vector
potential the Bethe wave functions exhibit boundary accumulation when the
ends are open~\cite{n_mao2023}. Oppositely directed
tunneling also occurs in cavity-assisted ladder models~\cite{b7}; here it
is the phase of a bond loss that selects the direction.

\subsection{A common transfer matrix for the two species}
The opposite hopping directions have a simple algebraic origin. Take
the ordinary Weyl pair $(c_+,d_+)=(a,a^\dagger)$ for one species and its
contragredient representation $(c_-,d_-)=(b^\dagger,-b)$ for the other.
Both obey $[c_\sigma,d_\tau]=\delta_{\sigma\tau}$. Set $\alpha=t/g$ and
$\nu_j=n_j^a-n_j^b-1$. In a three-dimensional auxiliary space, with $P_3$
the permutation of two such spaces, the local Lax operator and rational
$R$ matrix are
\begin{gather}
L_j(u)=
\begin{pmatrix}
u-\nu_j&\sqrt\alpha\,a_j&\sqrt\alpha\,b_j^\dagger\\
\sqrt\alpha\,a_j^\dagger&-\alpha&0\\
\sqrt\alpha\,b_j&0&+\alpha
\end{pmatrix},
\qquad R(u)=u\I-P_3.
\label{eq:14}
\end{gather}
The Weyl commutators give the RLL identity
\begin{equation}
R_{12}(u-v)L_{1j}(u)L_{2j}(v)
=L_{2j}(v)L_{1j}(u)R_{12}(u-v),
\label{eq:15}
\end{equation}
so the transfer matrices
\begin{align}
\tau(u)&=\Tr_{\rm aux}[L_L(u)\cdots L_1(u)]\notag\\
&=u^L+C_1u^{L-1}+C_2u^{L-2}+\cdots
\label{eq:boson-transfer}
\end{align}
commute for different $u$. For $L\geq3$, the leading coefficients of
Eq.~\eqref{eq:boson-transfer} are
\begin{align}
C_1&=-\sum_j\nu_j,\notag\\
C_2&=\frac12\sum_{i\neq j}\nu_i\nu_j+
\alpha\sum_j(a_j^\dagger a_{j+1}+b_{j+1}^\dagger b_j).
\label{eq:16}
\end{align}
The target belongs to this commuting charge algebra:
\begin{align}
\frac g2(C_1^2+C_1-2C_2)+gN_a-2gN_b
=H_{\leftrightarrow}+gL.
\label{eq:17}
\end{align}
Since both particle numbers commute with $\tau(u)$, the periodic conditional
Hamiltonian is Yang--Baxter integrable as well.

\begin{figure}[tb]
\centering
\includegraphics[width=\columnwidth]{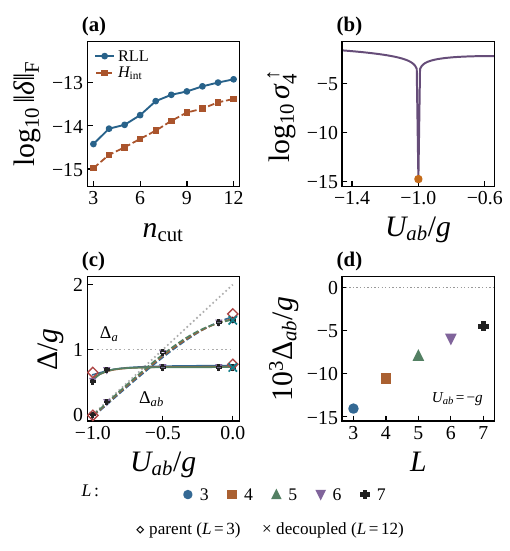}
\caption{Counterpropagating Bose--Hubbard mixture. (a) Projected RLL and
interaction residuals versus the local occupation cutoff. (b) Fourth
smallest singular value of the three-particle matching matrix.
Both ordinates in (a),(b) show base-ten logarithms. The extra
solution occurs at $U_{ab}=-g$ in this scan. (c) Single-component (solid)
and pair (dashed) curvatures at $t/g=0.08$, $N_a=N_b=L$. Lines join the
21-point scans for $L=3$--5; symbols give five sampled values for $L=6,7$.
Gray dotted lines denote the atomic limits, open diamonds the $L=3$
Hermitian parent, and crosses the decoupled $L=12$ endpoint.
(d) Signed pair curvature at $U_{ab}=-g$, enlarged by a factor of $10^3$.}
\label{fig:2}
\end{figure}

Equation~\eqref{eq:16} identifies the directed hopping of both species in
a single charge. The $\nu_j$ terms in Eq.~\eqref{eq:17} reproduce an onsite
interaction proportional to $(n_j^a-n_j^b)^2$: the same representation
that reverses the $b$ hopping fixes $U_{ab}=-g$. For a collision of two $a$
bosons and one $b$ boson, the coordinate-matching equations acquire a
nondiffractive solution at this coupling [Fig.~\ref{fig:2}(b)]. The extra
solution appears only at $U_{ab}/U_{aa}=-1$ in the displayed scan.
Appendix~\ref{app:A} gives the matching conditions and the finite-cutoff
checks of the oscillator identities.

\subsection{Spectrum of the particle-losing chain}
A jump removes one boson from each side of a density-matrix element:
recycling sends the ket and bra particle numbers $(M,M')$ to
$(M-1,M'-1)$. Conditional propagation preserves both. Ordering the
operator space by $M+M'$ thus makes the full Liouvillian block triangular
on every invariant finite-particle cutoff space. Its spectrum, including
algebraic multiplicities, consists of
\begin{equation}
\lambda_{mn}^{(M,M')}
=-i\left(\varepsilon_m^{(M)}-\varepsilon_n^{(M')*}\right).
\label{eq:22}
\end{equation}
Here $\varepsilon_m^{(M)}$ is a conditional energy in sector $M$,
including the loss shift $-itM$. This gives the two-component counterpart
of the loss grading in Ref.~\cite{b6}: collisions enter the diagonal
blocks through the conditional energies, while recycling connects
successive particle sectors. Evaluating all the $\varepsilon_m^{(M)}$
remains a separate spectral problem. In the auxiliary order $(-,0,+)$,
$L_j(u)$ is lower triangular on the Fock vacuum, with diagonal entries
$(\alpha,u+1,-\alpha)$. These vacuum weights give a starting point for
a nested Bethe ansatz.

\subsection{Energy cost of adding a pair}
At unit filling for each species, a particle and a pair probe different
directions of the onsite interaction. To connect the decoupled Mott
regime~\cite{b5,b10} to the coupled integrable point, fix $N_a=N_b=L$,
hold $t$ and $g$ fixed, and tune $U_{ab}$ from $0$ to $-g$.
Let $E_0(N_a,N_b)$ be the
target eigenvalue with smallest real part~\cite{b5}. For $t>0$, the
fixed-particle hopping graph is irreducible and all off-diagonal matrix
elements are nonpositive. Perron--Frobenius applied to $c\I-H$ with
sufficiently large real $c$ makes $E_0$ real and simple.
The canonical single-component and pair curvatures are
\begin{align}
\Delta_a={}&E_0(N_a+1,N_b)+E_0(N_a-1,N_b)-2E_0(N_a,N_b),\notag\\
\Delta_b={}&E_0(N_a,N_b+1)+E_0(N_a,N_b-1)-2E_0(N_a,N_b),\notag\\
\Delta_{ab}={}&E_0(N_a+1,N_b+1)+E_0(N_a-1,N_b-1)\notag\\
&-2E_0(N_a,N_b).
\label{eq:18}
\end{align}
In the atomic limit the local energy is
$E(n_a,n_b)=g[n_a(n_a-1)+n_b(n_b-1)]/2+U_{ab}n_an_b$.
For $g>0$ and $-g\leq U_{ab}\leq0$, minimization in the relevant
particle sectors gives the single-component curvatures
$\Delta_a^{(0)}=\Delta_b^{(0)}=g$ and the pair curvature
$\Delta_{ab}^{(0)}=2(g+U_{ab})$.

At $U_{ab}=0$, the components form independent unidirectional chains;
each has the unit-filling spectral Mott transition of Ref.~\cite{b5}
at $t_c/g\simeq0.17$. Attraction progressively lowers the atomic cost
of adding and removing a pair. This cost stays positive for
$-g<U_{ab}<0$ and vanishes at the coupled integrable point, where the
single-component curvatures still equal $g$. Indeed $E(m,m)=-gm$:
moving an $ab$ pair between sites with $n_j^a=n_j^b$ leaves the
total atomic energy unchanged. The soft direction is the
redistribution of pairs at fixed local population imbalance.

Finite hopping lifts the degeneracy. Figure~\ref{fig:2}(c) shows the
curvatures at $t/g=0.08$, with the atomic values and a reciprocal-parent
control shown separately. At the interacting point $U_{ab}=-g$,
Fig.~\ref{fig:2}(d) follows the signed pair curvature from $L=3$ to $L=7$.
At this coupling the single-component curvature stays positive. The pair
curvature, which measures the spacing between successive pair addition
energies, is negative at every size in panel (d) and decreases in
magnitude with $L$. Determining whether
the pair channel remains gapless at finite hopping requires the
thermodynamic behavior of these sector energies.

At one boson per site in total, the same attraction puts a neighboring
doublon--hole configuration below two singly occupied sites. The
second-order spin exchange in Appendix~\ref{app:A} consequently acts
within an excited manifold, in contrast to the ground-manifold exchange
of repulsive Mott insulators~\cite{b12}.

\subsection{Engineering counterpropagating bosons}
\label{sec:boson-realization}
\begin{figure}[tb]
\centering
\includegraphics[width=\columnwidth]{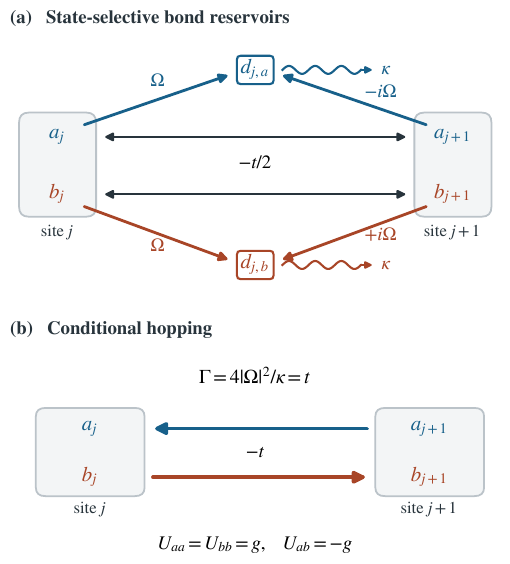}
\caption{Proposed bond-reservoir implementation for two internal states in
the same lattice wells. (a) The direct link has hopping $-t/2$. The two
paths to a given lossy auxiliary $d_{j,s}$ interfere coherently; the
auxiliaries for $a$ and $b$ decay through separate outputs. (b) Opposite
relative phases and $\Gamma=t$ select opposite hopping directions in the
conditional Hamiltonian. Atoms of the two species still interact on each
physical site.}
\label{fig:boson-realization}
\end{figure}

An atom can leave a bond through either endpoint. If both paths feed the
same decaying orbital, their relative phase controls which hopping amplitude
survives. Dissipative Aharonov--Bohm rings have demonstrated this
interference in a momentum-space lattice~\cite{n_gou2020}. The single-component proposal
of Ekman and Bergholtz~\cite{b6} uses a lossy orbital on each bond. For the
interacting mixture, consider two internal states sharing the same lattice
wells, with bare tunneling $-t/2$ for both states. The additional ingredients
are opposite phases in two state-selective loss channels and a tunable
attraction between atoms occupying the same well.

Figure~\ref{fig:boson-realization}(a) shows one bond. For each species
$s=a,b$, two phase-locked drives couple its endpoints to an auxiliary
orbital $d_{j,s}$, initially empty. Put $Q_{j,s}=s_j+e^{i\phi_s}s_{j+1}$. In the rotating
frame the resonant coupling is
\begin{equation}
H_{\rm aux}=\sum_{j,s}\bigl(\Omega d_{j,s}^{\dagger}Q_{j,s}
+\mathrm{H.c.}\bigr).
\label{eq:boson-aux-coupling}
\end{equation}
Atoms in each auxiliary escape to an independent reservoir at population
decay rate $\kappa$. For $\kappa$ large compared with the occupied-state
coupling matrix elements, the tunneling, and the local interaction shifts,
adiabatic elimination gives $J_{j,s}=\sqrt{\Gamma}\,Q_{j,s}$ to leading
order, with $\Gamma=4|\Omega|^2/\kappa$ and vanishing coherent shift on
resonance~\cite{b18}. The phases are measured relative to the real bare
hopping in Eq.~\eqref{eq:10}. At $\phi_a=-\pi/2$ and $\phi_b=+\pi/2$,
the left- and right-going amplitudes for $a$ are $-(t+\Gamma)/2$ and
$-(t-\Gamma)/2$; their order is reversed for $b$. Matching $\Gamma=t$
therefore gives the two jumps in Eq.~\eqref{eq:11} and the counterpropagating
bond in Fig.~\ref{fig:boson-realization}(b). The periodic model requires
the same coupling and phases on the closing bond.

A single isolated open bond can calibrate the cancellation before loading a
mixture. Prepare one atom at either endpoint, evolve for a time
$\kappa^{-1}\ll\tau\ll t^{-1}$, and retain shots with an atom on either
physical site. At the matched rate
only the leftward transfer of $a$ and the rightward transfer of $b$ remain.
With the phases held fixed, the short-time ratio of suppressed to allowed
transfer probabilities approaches $[(t-\Gamma)/(t+\Gamma)]^2$.
The interacting test also requires $U_{aa}=U_{bb}=g$ and $U_{ab}=-g$.
The choice of internal states, their Wannier overlaps, and the field near an
interspecies scattering resonance jointly set the three onsite energies.
A suitable mixture and confinement must meet both interaction conditions
while suppressing spin-changing and inelastic collisions. The three energies can be measured separately from the shifts of
same-state and mixed-state doublons relative to separated atoms. Tuning the
loss channels and the collision energies independently would allow the
directional hopping and the interaction balance to be tested in succession.
\section{Conclusion and outlook}\label{sec:conclusion}
We constructed a spin-1 chain whose isolated periodic bulk has no
finite-range conserved charge independent of the energy and the identity.
One common reservoir channel on each bond cancels the reverse spin
conversion. Products of the surviving moves along successive bonds then
give commuting charges of increasing range for the open chain. All
conditional energies coincide, yet the recycling term separates the decay
rates. The triangular Liouvillian gives the full spectrum and a dissipative
gap independent of chain length. The
reservoir preserves every superposition of blocked configurations because
they all give the same jump amplitude on a given bond. If the reservoir
can distinguish the local outputs, each bond on which two configurations
differ contributes equally to their dephasing, even though the conditional
Hamiltonian and its charges are unchanged.

For spin-$1/2$ the construction recovers the known Class-4 interaction.
In the Bose--Hubbard mixture, the phases of two bond-loss channels select
opposite hopping directions for species that still interact whenever
they meet. At $U_{aa}=U_{bb}=-U_{ab}$, a common transfer matrix on the
ring generates their interacting Hamiltonian up to a constant in each
fixed-number sector. Redistributing an $ab$ pair between sites then costs no
interaction energy in the atomic limit, while the single-particle gap for
either species remains finite. Each loss removes one particle, so at any
finite particle cutoff the Liouvillian spectrum follows from the
conditional spectra in fixed-number sectors.

Which blocked spin superpositions remain long lived if a small reverse
conversion amplitude survives on each bond? For the bosons, the open question is
whether the pair channel remains gapless at finite hopping in the
thermodynamic limit. A nested Bethe ansatz for the common transfer
matrix could address this by comparing the energies for adding and
removing one atom of each species. The proposed optical-lattice
geometry allows two independent tests: species-resolved tunneling on a
single bond would probe the directional cancellation, while onsite
spectroscopy would measure the interaction energies of like and unlike pairs.

\begin{acknowledgments}
This work was supported by the National Natural Science Foundation of China
(Grant No.~12304290 and No.~12674355) and the Beijing National Laboratory for Condensed Matter Physics (Grant No.~2025BNLCMPKF017). L.P. acknowledges support from the Fundamental Research Funds for the Central Universities (010-63263115). Generative AI tools (OpenAI Codex) assisted with manuscript editing and code development.
\end{acknowledgments}
\appendix
\section{Jump factors}\label{app:B}
\subsection{Two-qutrit jump}
Let $P_r$ reverse the nine-dimensional bond basis and let $P_rKP_r=L_cL_c^\dagger$ be the ordinary
lower Cholesky decomposition. The factor $J=P_rL_c^\dagger P_r$ is lower
triangular and obeys $J^\dagger J=K$. With the diagonal part $D$ specified at
Eq.~\eqref{eq:47}, the entries of $J-D$ are as follows; rows denote output
states and columns denote input states.
\begin{center}
\begingroup\small\setlength{\tabcolsep}{3pt}
\begin{tabular}{c@{\quad}rrrr}
\toprule
& $00$ & $01$ & $10$ & $11$\\
\midrule
$10$ & $0.1029857301$ & $0.6620511221$ & $0$ & $0$\\
$11$ & $-0.1507556723$ & $0.4522670169$ & $-0.6030226892$ & $0$\\
$21$ & $-i/2$ & $-i/2$ & $i$ & $i/2$\\
$22$ & $-i/2$ & $i/2$ & $-i/2$ & $-i$\\
\bottomrule
\end{tabular}
\endgroup
\end{center}
All omitted entries of $J-D$ vanish. Decimal entries are rounded;
the exact amplitudes are given with Eq.~\eqref{eq:47}.
Direct multiplication using the unrounded factor gives a factorization residual
$\|J^\dagger J-K\|=4.4\times10^{-16}$ in double precision.

\section{Yang--Baxter structure and open-chain charges}\label{app:C}
\subsection{Local braid identity}
For $\check R(u)=\I+uC$, the braided Yang--Baxter equation is
\begin{align}
\check R_{12}(u)\check R_{23}(u+v)\check R_{12}(v)\notag\\
=\check R_{23}(v)\check R_{12}(u+v)\check R_{23}(u).
\label{eq:54}
\end{align}
Expanding the left side of Eq.~\eqref{eq:54} yields
\begin{align}
\I+(u+v)(C_{12}+C_{23})+u(u+v)C_{12}C_{23}\notag\\
+v(u+v)C_{23}C_{12}+uvC_{12}^2+uv(u+v)C_{12}C_{23}C_{12}.\label{eq:C1}
\end{align}
The right side has the same first four terms, followed by
$uvC_{23}^2+uv(u+v)C_{23}C_{12}C_{23}$. Equation~\eqref{eq:52} therefore proves the identity for
arbitrary complex $u,v$. Integer arithmetic gives exactly zero for the three defining residuals;
a spectral test at $u=0.37,v=-0.61$ gives a floating-point residual below $10^{-14}$.

\subsection{Open-chain commuting local charges}
Put $C_j=C_{j,j+1}$ for $1\leq j\leq L-1$ and $H_C^{\rm OBC}=\sum_{j=1}^{L-1}C_j$.
The support of Eq.~\eqref{eq:42} gives three elementary rules. Every nonzero action of $C_j$
creates $\ket2$ at site $j$. A later generator touching that site annihilates
the term, while a disjoint generator leaves it unchanged.
Consequently any word containing the same generator twice vanishes. Moreover,
\begin{equation}
 C_jC_{j+1}=0,\qquad [C_i,C_j]=0\quad (|i-j|>1).
 \label{rev:oriented-relations}
\end{equation}
For an interval of bonds $I=[a,b]$, define $D_I=C_bC_{b-1}\cdots C_a$, and set
\begin{equation}
 S_k=\sum_{a=1}^{L-k}D_{[a,a+k-1]},\qquad 1\leq k\leq L-1.
 \label{rev:open-charges}
\end{equation}
Each term in $S_k$ acts on $k+1$ consecutive sites. We prove directly that
\begin{equation}
 [S_k,S_\ell]=0\quad\hbox{for all }L,k,\ell.
 \label{rev:open-commutativity}
\end{equation}
If two bond intervals overlap, both products vanish because a generator is repeated.
If their bond distance exceeds one, their products commute. For adjacent disjoint
intervals, multiplying the right interval by the left interval gives the operator
on their union, whereas multiplying them in the reverse order gives zero by
Eq.~\eqref{rev:oriented-relations}. In $[S_k,S_\ell]$, each union interval of length
$k+\ell$ therefore occurs once with coefficient $+1$ and once with coefficient $-1$:
the first partition places the $k$-interval on the right, and the second places it
on the left. These contributions cancel, including at both free boundaries.
This proves Eq.~\eqref{rev:open-commutativity} for every $L$ and establishes
the commuting hierarchy stated in Eq.~\eqref{eq:charges-main}.

The charges are nonzero and linearly independent. In the product basis,
\begin{equation}
 \bra{2^k1\,0^{L-k-1}}S_\ell\ket{0^L}
 =-2^{k-1}\delta_{k\ell}.
 \label{rev:independence-witness}
\end{equation}
For $\ell=k$, the first conversion has amplitude $W_{21,00}=-1$ and each subsequent
conversion uses $W_{21,10}=2$. A shorter interval cannot change both endpoints of
the displayed string, while a longer interval creates more than $k$ sites in
$\ket2$. Equation~\eqref{rev:independence-witness} proves independence without a
finite-size rank extrapolation. Since $S_1=H_C^{\rm OBC}$, these are $L-1$ commuting
local charges. They also commute with
$H_{\rm eff}^{\rm OBC}=H_C^{\rm OBC}-2i(L-1)\I$.

The same result has a Yang--Baxter generating-family formulation. Set
$X_j(u)=\I+uC_j$. The local relations give
$X_j(u)X_j(v)=X_j(u+v)$, distant commutativity, and the additive braided
Yang--Baxter relation. Corollary~5.4 of Fomin and Kirillov
\cite{b27} then applies directly to the open ordered product
\begin{align}
 T_{\rm OBC}(u)&=X_{L-1}(u)\cdots X_1(u),\notag\\
 [T_{\rm OBC}(u),T_{\rm OBC}(v)]&=0,\qquad
 T_{\rm OBC}'(0)=H_C^{\rm OBC}.
 \label{rev:open-family}
\end{align}
The product ends at bonds $1$ and $L-1$; it contains no bond between $L$ and $1$.
This supplies the commuting family for the open conditional chain used in the
Lindblad calculation.

\subsection{Additive charges and comparison under onsite similarities}\label{app:comparison}
The complex additive commutant of the local target is
\begin{align}
\mathfrak c_{\mathbb C}(C)
&=\{Q:[C,Q\otimes\I+\I\otimes Q]=0\}\notag\\
&=\operatorname{span}_{\mathbb C}\{\I,\ket2\bra0\}.
\label{eq:complex-commutant}
\end{align}
The map $Q\mapsto[C,Q\otimes\I+\I\otimes Q]$
is an $81\times9$ integer matrix in the matrix-unit basis. Its rank modulo
$1000003$ is seven. The two independent exact kernel vectors $\I$ and
$E_{20}$ supply the matching rational upper bound, proving
Eq.~\eqref{eq:complex-commutant} over $\mathbb C$. A nonscalar matrix
$x\I+yE_{20}$ has a nontrivial Jordan block. Since a uniform onsite
similarity conjugates both $C$ and its additive commutant, absence of
nonscalar diagonalizable elements is preserved.
In particular, no uniform invertible onsite similarity maps this density to
the number-conserving eclectic or hypereclectic densities of Ref.~\cite{b23},
including their one-coupling limits. The rank-one qutrit supersymmetric lifts
displayed in Refs.~\cite{b24,b25} have a nonscalar central support projector
$qq^\dagger+q^\dagger q$. This also distinguishes those densities, and
nontrivial additive-charge-conserving qutrit solutions~\cite{b26}, within the
same similarity class. The local descriptors are
\begin{align}
\rank C&=2,&C^2&=0,\notag\\
\OSR(C)&=2,&\dim\mathfrak c_{\mathrm{Herm}}(C)&=1.
\label{eq:58}
\end{align}
This comparison concerns uniform onsite similarities. Under a nonunitary
similarity Hermiticity can change, which is why the invariant uses the
complex additive commutant. Staggered similarities, twists, and
higher-dimensional representations lie outside this comparison.

\section{Periodic-bulk local-charge obstruction}\label{app:D}
The local density is decomposed into its connected traceless interaction
and one-site terms,
\begin{equation}
h=h^{(2)}+a\otimes\I+\I\otimes b,\qquad h^{(1)}=a+b.
\label{eq:37}
\end{equation}
We apply Hokkyo's theorem~\cite{b13} to the periodic completion of the
Hamiltonian. The required conditions are injectivity of both one-site
commutator maps of $h^{(2)}$, a one-dimensional zero-momentum space
$\mathcal B_{\leq}^{(2,0)}$ of two-site leading densities whose commutators
remain at most two-local, and absence of a strictly three-local density
whose commutator is at most two-local. Under these conditions, Theorem~1
of Ref.~\cite{b13} excludes strictly $k$-local conserved quantities for
$3\leq k\leq L/2$. Every fixed finite range is covered for sufficiently
large $L\geq2k$. This is a periodic-bulk statement; the spectral and
dynamical calculations use open boundaries.

We use the integer traceless one-site basis
\begin{equation}
\{E_{ab}:a\neq b\}\cup\{\operatorname{diag}(1,-1,0),\operatorname{diag}(0,1,-1)\}.\label{eq:D1}
\end{equation}
All entries of $6h^{(2)}$ and $6h^{(1)}$ are integers. Exact elimination
over $\mathbb F_{1000003}$ gives
\begin{align}
\rank M_{\mathrm{inj},L}&=\rank M_{\mathrm{inj},R}=8,\notag\\
\rank M_{2,0}&=63\quad(64\text{ columns}).
\label{eq:38}
\end{align}
Thus both one-site maps are injective and the two-site leading space has
dimension one, generated by $h^{(2)}$.
For each two-site basis matrix $q$, the highest-support map entering Hokkyo's Eq.~(77) is
\begin{equation}
\Phi_z(q)=[q_{12},h_{23}^{(2)}]-z[h_{12}^{(2)},q_{23}].\label{eq:D2}
\end{equation}
At $z=1$, its $729\times64$ matrix has modular rank 63. To exclude all $z\neq1$, we solve the
enlarged equation $[q_{12},h_{23}^{(2)}]=[h_{12}^{(2)},y_{23}]$ with independent $q,y$. Its
$729\times128$ matrix has rank 127 and the unique null vector has $q=y=h^{(2)}$, proving that
the only allowed value is $z=1$.

For the three-site step, define periodic translation sums
\begin{equation}
Q_b=\sum_j[h_{j,j+1}^{(2)},h_{j+1,j+2}^{(2)}],\qquad
Q_2=\sum_jq_{j,j+1}.\label{eq:D3}
\end{equation}
The only possible three-site leading density is proportional to
$[h_{12}^{(2)},h_{23}^{(2)}]$. Its arbitrary two-site correction would have
to satisfy
\begin{equation}
[Q_b+Q_2,H_0]=Y_2+Y_1.
\label{eq:39}
\end{equation}
On the smallest periodic chain without support ambiguities, $L=6$, this
is an integer system with 531441 rows and 136 unknowns: 64 coefficients
for $q$, 64 for $Y_2$, and 8 for $Y_1$. Sparse elimination over
$\mathbb F_{1000003}$ gives
\begin{equation}
\rank M_3=135,\qquad\rank(M_3\mid b)=136.
\label{eq:40}
\end{equation}
The modular ranks supply lower bounds on the rational ranks. The exact integer kernels
$h^{(2)}$ and $(h^{(2)},h^{(2)})$ provide the matching upper bounds for the 64- and 128-column
two-site systems. The exact nonzero homogeneous
solution obtained by setting $Q_2=\sum_jh_{j,j+1}^{(2)}$, with
$Y_2=[Q_2,\sum_jh_j^{(1)}]$ and $Y_1=0$, supplies the upper bound
$\rank_{\mathbb Q}M_3\leq135$. Hence the coefficient rank is exactly 135, whereas the augmented
rank is at least 136. The linear system is therefore inconsistent over $\mathbb Q$. Because its
coefficients are rational, it is also inconsistent over $\mathbb R$ and $\mathbb C$.
The fingerprint argument of Appendix~\ref{app:symmetry} also applies to
periodic connected bonds: a conserved site-dependent one-site sum
$Q_1=\sum_jq_j$ must obey
$[h^{(2)},q_j\otimes\I+\I\otimes q_{j+1}]=0$ on every bond, whose
partial-trace fingerprints force every $q_j$ to be scalar.
Since the two-site leading space is generated by the interaction density,
subtracting a multiple of the Hamiltonian leaves such a one-site sum.
Thus only the Hamiltonian and scalar identity remain in the stated
local-charge class.

\section{Triangular spectrum and steady-space proof}\label{app:E}
In lexicographic order, a bond embedding of a lower-triangular operator remains lower triangular
for open boundaries. Kronecker products and sums preserve this property, proving triangularity of
the column-vectorized generator in Eq.~\eqref{eq:5}. The diagonal of $\Heff$ is $-2i(L-1)$, while the recycling diagonal on
$\ket{\bm a}\bra{\bm b}$ is $\sum_jd_{a_ja_{j+1}}d_{b_jb_{j+1}}^*$. Their sum gives Eq.~\eqref{eq:60}.

To count allowed strings, let $x_L$ and $y_L$ count strings ending in $\ket2$ and in one of the
two low states. Then $x_L=x_{L-1}+y_{L-1}$ and $y_L=2x_{L-1}$. Eliminating $x_L$ gives
\begin{equation}
N_{\mathrm{DFS}}(L)=N_{\mathrm{DFS}}(L-1)+2N_{\mathrm{DFS}}(L-2),
\label{eq:65}
\end{equation}
with $N_{\mathrm{DFS}}(1)=3$ and $N_{\mathrm{DFS}}(2)=5$, leading to
Eq.~\eqref{eq:66}. Every term in the triangular spectrum is nonpositive
and vanishes only when both bond amplitudes equal two. The algebraic zero
multiplicity is therefore $N_{\mathrm{DFS}}^2$. The same number of
explicit stationary coherences proves semisimplicity at zero and
exhausts the stationary operator space.

\section{Numerical methods and reproducibility}\label{app:F}
The reproducibility package contains the spectra, time-dependent
observables, and convergence records used to construct the figures.

The bosonic algebra is tested by projected finite-Fock RLL residuals and
three-particle coordinate matching (Appendix~\ref{app:A}). For the energy
curvatures on $L=3$--6 rings, a sparse matrix-free solver acts directly on fixed $(N_a,N_b)$
sectors of the periodic counterpropagating target.
At $g=1$, $t=0.08$, $N_a=N_b=L$, the scan uses 21 uniformly spaced
$U_{ab}/g$ values in $[-1,0]$ for each $L=3,4,5$, and
$-1,-0.9,-0.5,-0.1,0$ for $L=6$. Seven particle sectors per point give
476 eigenpairs for $L=3$--6; the largest of these sectors has dimension 627264. These
eigenpairs meet the solver convergence check, with relative residuals below
$2.8\times10^{-15}$. The residual tests the eigenpair equation. We identify
the lowest-real-part branch using the matrix sign structure and compare it
with independent solutions in small sectors.

The $L=7$ calculation samples the same five coupling values used for
$L=6$. At finite hopping the positive ground-state vector is
invariant under simultaneous translation of the two species: translating
it gives a positive eigenvector with the same eigenvalue, and uniqueness
fixes the translation eigenvalue to one. We therefore represent its
amplitudes on translation orbits. Reflection followed by species exchange
relates the two single-component addition sectors and likewise the two
removal sectors. Five independent energies suffice to form both canonical
curvatures at each coupled point. In the largest sector, $(N_a,N_b)=(8,8)$,
the $9\,018\,009$ occupation states reduce to $1\,288\,287$ orbits.
We check the residual in the physical Hilbert norm, weighting each orbit
by its length. The largest absolute residual for a normalized $L=7$
state over the four coupled points is below $2\times10^{-11}g$. At
$U_{ab}=-g$ the five $L=6$ sector energies agree with the independent
full-sector calculation to within $6\times10^{-13}g$. At the decoupled endpoint, sums of single-species
energies give a separate check up to $L=12$.
The two-point $L=3$ reciprocal-mother control is plotted separately.
At $U_{ab}=-g$, the target pair curvatures in units of $g$ are
\begin{center}
\begin{tabular}{cr}
\toprule
$L$&$\Delta_{ab}/g$\\\midrule
3&$-0.01404735$\\
4&$-0.01054005$\\
5&$-0.00791710$\\
6&$-0.00596355$\\
7&$-0.00449719$\\\bottomrule
\end{tabular}
\end{center}
The $L=3$ mother gives $\Delta_a/g=0.64792999$ and
$\Delta_{ab}/g=-0.01099474$, compared with $0.61384583$ and
$-0.01404735$ for the target. We retain the signs of the second differences
in comparing the finite-hopping target, its Hermitian parent, and the atomic
reference. Locating a thermodynamic phase boundary would require the
size dependence beyond these chains.

The qubit calculation evaluates the Class-4 braid identity at
$(u,v)=(0.17,-0.31)$ and on a $31\times37$ spectral-parameter grid,
reconstructing the positive loss kernel at each parameter point.
For the qutrit mother, full real-symmetric diagonalization is used for
open chains $L=5,6,7,8$. After excluding exact zero modes at tolerance
$10^{-7}$, the full positive-level counts are 118, 363, 1096, and 3298.
Their central 80\% windows contain 95, 290, 877, and 2639 levels, giving
93, 288, 875, and 2637 adjacent-gap ratios. Both energy signs, central
40\%, 60\%, 80\% windows, and zero thresholds $10^{-8},10^{-7},10^{-6}$
give 72 consistency settings. The two-sided empirical-CDF distance is
\begin{equation*}
D_n=\max_i\left\{\frac{i}{n}-F(r_i),\,F(r_i)-\frac{i-1}{n}\right\}.
\end{equation*}
This distance quantifies agreement with the Atas or Poisson distribution
for the spectrum of the open chain. The proof of the bulk local-charge
obstruction uses the finite-field certificate in Appendix~\ref{app:D}.
The central-positive-window results are
\begin{center}
\begin{tabular}{crrr}
\toprule
$L$&positive levels&$\langle r\rangle$&$D_n(\mathrm{Atas})$\\\midrule
5&118&0.60720&0.14264\\
6&363&0.53931&0.04415\\
7&1096&0.52720&0.02528\\
8&3298&0.53476&0.01005\\\bottomrule
\end{tabular}
\end{center}
The positive-level column gives the full count before window selection.
The Atas surmise has mean $4-2\sqrt3=0.535898\ldots$, while the
numerical large-GOE reference is approximately $0.5307$; the Poisson
mean is $2\ln2-1=0.38629\ldots$~\cite{b16,b17}. The three zero
thresholds leave the reported ratios unchanged.

The local Reshetikhin defect of the mother is
\begin{equation}
\|D_R(h)\|_F=13.4908448,\qquad
\|D_R(h)\|_{\mathrm{rel}}=0.7284371,
\label{eq:41}
\end{equation}
where $\|D_R(h)\|_{\mathrm{rel}}=\|D_R(h)\|_F/\|h\|_F^3$
and $\|h\|_F=\sqrt7$. Its nonzero value excludes a regular
difference-form Yang--Baxter structure for this Hermitian density
under the Reshetikhin condition~\cite{b14,b15}; it is independent
of the stronger local-charge obstruction in Appendix~\ref{app:D}.

For the full qutrit Liouvillian at $L=3$, direct construction of the
$729\times729$ matrix gives zero upper-triangle norm, a trace-preservation
error of $1.5\times10^{-15}$, and precisely $11^2=121$ zero singular values.

Spin-1 dynamics uses Arnoldi exponential propagation of the complete
Lindblad generator without diagonalizing it. The Liouville dimensions
are 59049 and 531441 at $L=5,6$. We use open boundaries and unit overall
rate. Figure~\ref{fig:7}(a),(b) compares four product-state trajectories
on 61 common times in $0\leq t\leq10$. Panel (c) uses the site-resolved
occupations from the $L=6$ all-zero run at those times. The two $L=5$
cat-state trajectories in panel (d) have 201 sampling times and relative
phase $i$ as specified in the text.
Main/reference Arnoldi dimensions are 32/40 and maximum internal
steps are $0.05/0.025$; output times are distinct from internal steps.
The tolerance $10^{-13}$ controls the local Arnoldi-breakdown test.
We assess propagation accuracy by comparison with the reference integration.
The largest main/reference final-density Frobenius difference
for the paired product-state runs is $1.3\times10^{-15}$.
No trace renormalization, probability clipping, or Hermitian projection
is applied. Trace and probability bounds are checked at every saved time;
positivity is checked by diagonalizing the density matrix at $t=0,2.5,5,7.5,10$.
The maximum trace error is $3.2\times10^{-15}$.
The product-state endpoint values at $t=10$ are
\begin{center}
\begin{tabular}{clrr}
\toprule
$L$&initial state&$\langle A\rangle$&$\langle P_{\mathrm{DFS}}\rangle$\\\midrule
5&$\ket{00000}$&$2.39\times10^{-3}$&0.9907\\
5&$\ket{01010}$&$4.87\times10^{-4}$&0.9981\\
6&$\ket{000000}$&0.0023930819&0.9882166555\\
6&$\ket{010101}$&0.0004422573&0.9977959159\\\bottomrule
\end{tabular}
\end{center}
The active cat $(\ket{00000}+i\ket{11111})/\sqrt2$ has coherence
$5.3\times10^{-10}$ at $t=10$, whereas the coherence of
$(\ket{20202}+i\ket{21212})/\sqrt2$ remains $1/2$.

For Fig.~\ref{fig:8} we apply the physical bond generators directly
to vectors in the $9^L$-dimensional operator space, using the open-chain
Hamiltonian and the common and resolved jumps. For the chosen cat states
the matrix-free Arnoldi iteration closes after one or two basis vectors
without a prior projection onto the blocked subspace. We propagate
$L=2,4,6,7$ at $\eta=0,0.01,0.025,0.05,0.1,0.15,0.2,1$ on 81 times
from $0$ to $2$. Fitting $\ln(2|\rho_{\bm a\bm b}|)$ on the interval
$2|\rho_{\bm a\bm b}|\geq e^{-6}$ gives the rates.
Over these runs the full-density Frobenius discrepancy from
Eq.~\eqref{eq:output-coherence} is below $1.3\times10^{-15}$,
and the largest rate discrepancy is below $3.2\times10^{-14}$.
The active product states in Fig.~\ref{fig:7} require a larger Krylov basis.

\section{Resolved-output spectrum and gap}\label{app:resolved}
Multiplication by a diagonal projector preserves lower triangularity. The
recycling diagonal on $\ket{\bm a}\bra{\bm b}$ is therefore
$\sum_j[1-\eta+\eta\delta_{\alpha_j,\beta_j}]d_{\alpha_j}d_{\beta_j}$,
where $\alpha_j=(a_j,a_{j+1})$ and $\beta_j=(b_j,b_{j+1})$.
Together with the unchanged conditional term, this gives the full spectrum
\begin{equation}
\lambda^{(\eta)}_{\bm a,\bm b}
=\sum_{j=1}^{L-1}\left[
(1-\eta+\eta\delta_{\alpha_j,\beta_j})d_{\alpha_j}d_{\beta_j}-4
\right].
\label{eq:resolved-spectrum}
\end{equation}
All summands are nonpositive.
For $\eta>0$, a zero summand requires equal inactive local strings. Equality
on every bond of a connected open chain with $L\geq2$ implies $\bm a=\bm b$.
The $N_{\rm DFS}$ allowed-state projectors are explicit stationary operators
and exhaust the diagonal zero multiplicity, proving Eq.~\eqref{eq:resolved-kernel}
and semisimplicity of the zero eigenvalue.

The nonzero spectral gap of this family is
\begin{equation}
\Delta_{\Li}(\eta)=\min\left(4\eta,\frac8{15}\right),
\qquad 0<\eta\leq1,\quad L\geq2.
\label{eq:resolved-gap}
\end{equation}
Indeed, a bond with different local strings contributes a decay rate at least
$4-4(1-\eta)=4\eta$. A matching active bond contributes at least
$4-d_{00}^2=8/15$, since $d_{00}^2=52/15$ is the largest active diagonal
square. Every nonzero eigenvalue contains at least one of these contributions.
The bounds are attained by two allowed strings differing only at an endpoint,
or by identical strings with a single $00$ bond and all remaining sites in
$\ket2$, respectively.

At $\eta=0$ the coherences belong to the kernel and the first nonzero
decay rate is given by Eq.~\eqref{eq:63}. As $\eta\to0^+$ the gap in
Eq.~\eqref{eq:resolved-gap} follows the coherence modes into the kernel.
For a DFS cat with $m>0$, taking $t\to\infty$ first removes the coherence,
whereas taking $\eta\to0^+$ first preserves it. The change of stationary
space therefore occurs already on every finite chain with $L\geq2$,
independently of any thermodynamic phase transition.

\begingroup
\section{Onsite and reflection symmetries of the open mother}\label{app:symmetry}
Put $F=6h^{(2)}$. A product symmetry $\bigotimes_jU_j$ preserves the uniquely
defined connected bond term on every bond, so each $U_j\otimes U_{j+1}$ preserves
$F$. Define $A_r=\Tr_2F^r$ and $B_r=\Tr_1F^r$. Partial tracing gives
\begin{align}
 [U_j,A_2]&=[U_j,A_3]=0,\notag\\
 [U_{j+1},B_2]&=[U_{j+1},B_4]=0.\label{eq:symmetry-fingerprints}
\end{align}
The exact integer fingerprints are
{\setlength{\arraycolsep}{3pt}
\begin{align*}
 A_2&=\begin{pmatrix}33&-33&0\\-33&87&0\\0&0&120\end{pmatrix},\\
 A_3&=\begin{pmatrix}0&0&-93\\0&0&147\\-93&147&0\end{pmatrix},\\
 B_2&=\begin{pmatrix}67&18&0\\18&106&-18\\0&-18&67\end{pmatrix},\\
 B_4&=\begin{pmatrix}4292&2295&486\\2295&7940&-2493\\486&-2493&5084\end{pmatrix}.
\end{align*}
}
Each joint commutant has rank eight in nine complex matrix entries. Nonzero
integer eight-minors have determinants $-32543665449768$ and $-23920661090304$,
while the identity is an exact kernel vector. Thus every $U_j$ is scalar;
this includes staggered and nondiagonal onsite products.

Reflection exchanges the one-site endpoint contributions. Here
\begin{equation}
 \Tr_2(2h)=\begin{pmatrix}0&0&-1\\0&0&1\\-1&1&0\end{pmatrix},
 \qquad \Tr_1(2h)=0.
\end{equation}
The zero/nonzero distinction is preserved by arbitrary onsite-unitary
conjugation, excluding reflection with such a dressing. Complex conjugation
remains an antiunitary symmetry with square $+1$.
For $L=5,6,7,8$, graph analysis finds one nontrivial block of dimension $3^L-3$
and three isolated configurations. Three further zero modes occur in the large
block. All six zero levels are excluded before selecting positive/negative
windows; the retained nonzero levels are numerically nondegenerate.
\endgroup

\section{A common-output Raman construction}\label{app:implementation}
The effective-operator formalism of Ref.~\cite{b18} realizes the same
bond jump after eliminating a rapidly decaying auxiliary manifold.
Introduce nine auxiliary states $\ket{e_\alpha}$ corresponding to the
ground-bond states $\ket\alpha$, and couple all decay paths to one
indistinguishable environmental mode:
\begin{equation}
L_e=\sqrt\kappa\sum_{\alpha=1}^9\ket\alpha\bra{e_\alpha}.
\label{eq:55}
\end{equation}
Then $L_e^\dagger L_e=\kappa\I_e$. Phase-locked drives encode the
desired ground-manifold matrix elements through
\begin{equation}
V_+=\Omega\sum_{\alpha,\beta}J_{\alpha\beta}
\ket{e_\alpha}\bra\beta,\qquad V_-=V_+^\dagger.
\label{eq:56}
\end{equation}
On resonance, $H_{\mathrm{NH}}=-i\kappa\I_e/2$ and elimination gives
\begin{equation}
J_{\mathrm{eff}}=L_eH_{\mathrm{NH}}^{-1}V_+
=\frac{2i\Omega}{\sqrt\kappa}J.
\label{eq:57}
\end{equation}
The coherent elimination shift vanishes because
$H_{\mathrm{NH}}^{-1}+H_{\mathrm{NH}}^{-1\dagger}=0$.
Writing $\Gamma=4|\Omega|^2/\kappa$, the coherent mother must be
$\Gamma\sum_jh_j$ to reproduce the same dimensionless conditional
target. The approximation is controlled by
$|\Omega|\|J\|/\kappa\ll1$.

In this proposal several phase-locked Raman amplitudes feed one
indistinguishable decay channel. Their relative phases and magnitudes
set the individual matrix elements of the correlated two-qutrit jump.
Gate-and-reset engineering in trapped ions~\cite{b19} and cold-atom
reservoir engineering~\cite{b20} provide precedents for engineered
dissipation; implementing this specific jump requires control of all
the indicated paths.

\section{Bosonic RLL identity and finite-size checks}\label{app:A}
For the rational auxiliary matrix $R(u)=u\I-P_3$, substituting Eq.~\eqref{eq:14} into
\begin{equation}
R_{12}(u-v)L_{1j}(u)L_{2j}(v)=L_{2j}(v)L_{1j}(u)R_{12}(u-v)\label{eq:A1}
\end{equation}
reduces every entry to
\begin{equation}
[a_j,a_j^\dagger]=1,\quad[b_j^\dagger,-b_j]=1,\quad[a_j,b_j]=[a_j,b_j^\dagger]=0.\label{eq:A2}
\end{equation}
The second relation is the contragredient Weyl representation. RLL is an algebraic identity and
does not require $d_-=c_-^\dagger$.

The supplied verification program truncates each local oscillator at $0\leq n_a,n_b\leq n_{\mathrm{cut}}$
and projects the residual away from the two artificial upper Fock boundaries. At $n_{\mathrm{cut}}=6$,
the interior Frobenius RLL residual is $1.7642\times10^{-14}$, while the three Weyl-commutator residuals
are at most $5.5467\times10^{-15}$. These checks use $\alpha=0.8$, $g=0.7$,
$u=0.37$, and $v=-0.61$, with $n_a,n_b<n_{\mathrm{cut}}$ in the interior projection. The charge-to-Hamiltonian identity is checked independently.
Its local interaction content follows from
\begin{align}
\frac g2\nu_j(\nu_j-1)={}&\frac g2[n_j^a(n_j^a-1)+n_j^b(n_j^b-1)]\notag\\
&-gn_j^a n_j^b-gn_j^a+2gn_j^b+g,\label{eq:A3}
\end{align}
so the remaining terms in Eq.~\eqref{eq:17} are only conserved linear charges and a constant.
The finite-cutoff interaction-mapping residual is $4.9861\times10^{-15}$.

\paragraph{Explicit three-particle coordinate matching.}
We use two identical $a$ bosons at coordinates $x,y$ and one $b$ boson at $z$,
with $x\leq y$ imposed by bosonic exchange symmetry. In first quantization,
define $V_{xyz}=U_{aa}\delta_{xy}+U_{ab}(\delta_{xz}+\delta_{yz})$. The target satisfies
\begin{align*}
E\psi(x,y,z)={}&-t\bigl[\psi(x+1,y,z)+\psi(x,y+1,z)\bigr]\\
&-t\psi(x,y,z-1)+V_{xyz}\psi(x,y,z).
\end{align*}
The shifts follow directly from $a_j^\dagger a_{j+1}$ and
$b_{j+1}^\dagger b_j$. Define the eight weak-order sectors
\begin{gather*}
x<y<z,\quad x<z<y,\quad z<x<y,\\
x=y<z,\quad z<x=y,\quad x=z<y,\\
x<y=z,\quad x=y=z.
\end{gather*}
Within sector $r$, the nondiffractive ansatz is
\begin{gather*}
\psi(x,y,z)=A_r z_1^x z_2^y w^z+B_r z_2^x z_1^y w^z,\\
E=-t(z_1+z_2+w^{-1}).
\end{gather*}
Amplitudes on coincidence sectors are retained independently; no two-body
factorization or triple-coincidence equation is assumed in advance.
Writing $F(x,y,z)$ for the row vector of these 16 basis functions gives
the matching row
\begin{align*}
M_{x,y,z}={}&(E-V_{xyz})F(x,y,z)\\
&+t\bigl[F(x+1,y,z)+F(x,y+1,z)\bigr]\\
&+tF(x,y,z-1).
\end{align*}
Coordinates are sorted again before each evaluation of $F$, implementing
the exchange symmetry of the two $a$ particles.

We set $t=U_{aa}=1$ and
\begin{equation*}
(z_1,z_2,w)=(e^{0.37i},e^{1.13i},e^{-0.79i}),
\end{equation*}
and stack rows at $x=0$, $y=0,\ldots,5$ and $z=-5,\ldots,10$.
This gives a $96\times16$ matrix. The three coincidence sectors with
$x=y$ have identical $A_r$ and $B_r$ columns and therefore supply three
parameter-independent null vectors. At the threshold $10^{-10}$ the
nullity is three at every sampled point in
$U_{ab}/U_{aa}\in[-1.45,-0.55]$ except $-1$, where it is four.
The fourth singular value from the bottom is
$1.6150\times10^{-15}$ at $U_{ab}/U_{aa}=-1$.

Deleting one repeated column in each of those three sectors yields a
$96\times13$ physical matching matrix. Its nonzero null vector at the
coupled point is checked independently on 225 rows with $x=0$,
$y=0,\ldots,8$, $z=-8,\ldots,16$; the residual norm is
$3.8010\times10^{-15}$ for a unit-norm coefficient vector.
The decoupled endpoint $U_{ab}=0$ also has nullity four, consistently
with two independent single-component chains; it lies outside the plotted
interval. The stored data include all matching matrices, singular values,
coordinates and coefficients.

\subsection{Formal spin projection and its limitation}\label{app:boson-projection}
The attraction at $U_{ab}=-g$ prevents the ground-state spin description
obtained from repulsive Mott states~\cite{b12}. On one bond,
$\ket{b_j a_{j+1}}$ can reach $\ket{a_j b_{j+1}}$ through either
intermediate doublon--hole configuration, $\ket{(ab)_j0_{j+1}}$ or
$\ket{0_j(ab)_{j+1}}$. Their energy is $-g$, below the energy zero of
the singly occupied manifold. The denominator
$E_{\rm single}-E_{ab}=g$ gives $t^2/g$ per path and the formal exchange
\begin{equation}
H_{\rm spin}^{(2)}
=\frac{2t^2}{g}\sum_j\ket{a_jb_{j+1}}\bra{b_ja_{j+1}}.
\label{eq:24}
\end{equation}
For equal hopping directions the exchange path is absent. Opposite
directions yield the familiar one-way spin-$1/2$ exchange in
Eq.~\eqref{eq:24}. The doublon--hole states lie below the singly occupied
manifold, so the projection describes a formal exchange process in an
excited manifold and cannot furnish a ground-state spin Hamiltonian.

\section{Known Class-4 spin-1/2 benchmark}\label{sec:qubit}\label{app:qubit}
\begin{figure*}[t]
\includegraphics[width=\textwidth]{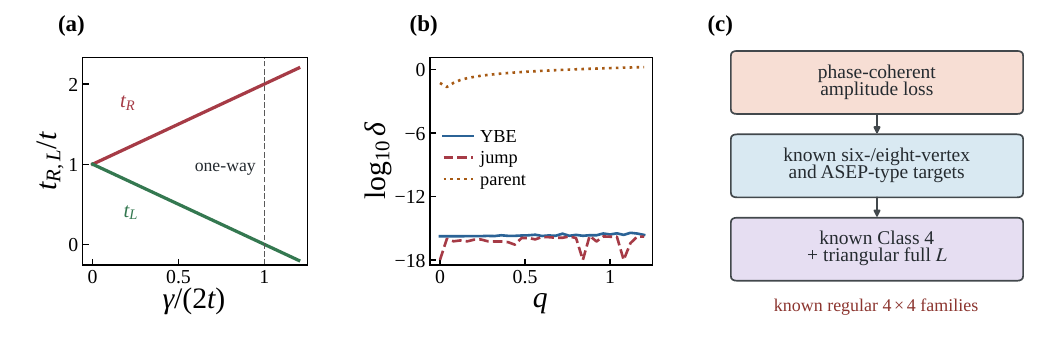}
\caption{\textbf{Directed spin processes and a solvable qubit example.}
(a) Phase-coherent loss can generate directed exchange.
(b) The Class-4 construction satisfies the local Yang--Baxter identity and
loss-kernel factorization, while the tested Hermitian density fails the
regular difference-form Reshetikhin condition. The ordinate is
$\log_{10}\delta$; for display, the braid and reconstruction residuals
are bounded below by $10^{-18}$.
(c) The displayed conditional bulk target is known. Its upper-triangular
bond jump separately yields the full open-chain Liouvillian spectrum in
Eq.~\eqref{eq:34}.}\label{fig:3}
\end{figure*}

Figure~\ref{fig:3} illustrates the construction for the known Class-4
spin-$1/2$ family~\cite{b8,b9}.

\subsection{Reciprocal couplings and the bond jump}
Consider spin-1/2 particles with the real bond Hamiltonian
\begin{align}
h_0={}&\sigma^z\otimes\sigma^z+\frac p4(\I\otimes\sigma^x+\sigma^x\otimes\I\notag\\
&+\sigma^x\otimes\sigma^z+\sigma^z\otimes\sigma^x)
+\frac q4(\sigma^x\otimes\sigma^x-\sigma^y\otimes\sigma^y),\label{eq:27}
\end{align}
where $p,q$ are real. The open-chain Hamiltonian is
\begin{equation}
H_{1/2,0}=\sum_{j=1}^{L-1}h_{0,j,j+1}.\label{eq:26}
\end{equation}
The Ising interaction is accompanied by transverse fields, conditional
spin flips, and reciprocal pair creation and annihilation.

In the basis $\{\ket{00},\ket{01},\ket{10},\ket{11}\}$, define
\begin{equation}
h_4=\begin{pmatrix}1&p&p&q\\0&-1&0&0\\0&0&-1&0\\0&0&0&1\end{pmatrix},\label{eq:25}
\end{equation}
so that $h_0=(h_4+h_4^\dagger)/2$. The required loss kernel is
\begin{equation}
K_4=\kappa\I_4+i(h_4-h_4^\dagger),\label{eq:28}
\end{equation}
with $\kappa>\kappa_{\min}\equiv-\lambda_{\min}[i(h_4-h_4^\dagger)]$.
Choose the upper-triangular Cholesky factor as the repeated bond jump,
\begin{equation}
J_{4,j}=\chol(K_4),\qquad J_{4,j}^\dagger J_{4,j}=K_4.\label{eq:29}
\end{equation}
The matrix elements of $J_4$ add coherently within a single two-spin jump.
An explicit factor is given in Sec.~\ref{app:qubit-jump}.

\subsection{Conditional directed dynamics}
Combining Eqs.~\eqref{eq:26} and~\eqref{eq:29} gives
\begin{equation}
H_{1/2,\mathrm{eff}}=\sum_{j=1}^{L-1}h_{4,j,j+1}
-\frac{i\kappa}{2}(L-1)\I.\label{eq:31}
\end{equation}
The off-diagonal entries now convert $\ket{01}$, $\ket{10}$, and $\ket{11}$
into $\ket{00}$.

The density $h_4$ belongs to the known Class-4 Yang--Baxter
family~\cite{b8,b9}. Its regular braided matrix is
\begin{equation}
\check R_4(u)=\begin{pmatrix}
e^u&p\sinh u&p\sinh u&e^u q\cosh u\sinh u\\
0&e^{-u}&0&0\\
0&0&e^{-u}&0\\
0&0&0&e^u
\end{pmatrix},\label{eq:32}
\end{equation}
with $\check R_4(0)=\I$ and $\check R'_4(0)=h_4$. This identifies the
conditional bulk density; a boundary-compatible transfer hierarchy for
the open qubit chain remains an independent problem. At the parameters
of Sec.~\ref{app:qubit-jump}, the reciprocal Hermitian density violates
the Reshetikhin condition for a regular difference-form $R$ matrix.

\subsection{Full open-chain relaxation spectrum}
Both $h_4$ and $J_4$ are upper triangular, as are their open-chain
embeddings in Eq.~\eqref{eq:5}. With $\epsilon=(1,-1,-1,1)$ the diagonal of $h_4$ and
$r_\alpha=(J_4)_{\alpha\alpha}>0$, the full Liouvillian has eigenvalues
\begin{align}
\lambda_{\bm a,\bm b}^{(1/2)}=\sum_{j=1}^{L-1}\big[&-i(\epsilon_{a_j a_{j+1}}
-\epsilon_{b_j b_{j+1}})-\kappa\notag\\
&+r_{a_j a_{j+1}}r_{b_j b_{j+1}}\big].\label{eq:34}
\end{align}
The differences of Ising energies give the oscillation frequencies, while the
diagonal jump amplitudes and the uniform loss shift fix the decay rates.

\subsection{Two-qubit jump and local diagnostics}\label{app:qubit-jump}
For the representative qubit parameters $p=0.4$ and $q=0.7$, the positivity
threshold in Eq.~\eqref{eq:28} is $\kappa_{\min}=0.9$. Choosing $\kappa=1$
in Eq.~\eqref{eq:29} gives
\begin{equation}
J_4=\begin{pmatrix}
1&0.4i&0.4i&0.7i\\
0&0.916515&-0.174574&-0.305505\\
0&0&0.899735&-0.370479\\
0&0&0&0.528594
\end{pmatrix}.\label{eq:30}
\end{equation}
These entries are rounded; the Cholesky definition specifies the exact
operator and is used in the calculations. At this point the numerical
Yang--Baxter residual is below $4.2\times10^{-16}$, and the reconstruction
of Eq.~\eqref{eq:31} agrees to machine precision. The optimized local
Reshetikhin defect of the Hermitian density, including Hermitian telescopic
gauges, is $0.9405$. The tested Hermitian density thus fails the
regular difference-form criterion even after optimizing over those gauges.
\bibliography{references}
\end{document}